\documentclass[sn-nature,Numbered,iicol]{sn-jnl}

\usepackage{graphicx}%
\usepackage{multirow}%
\usepackage{amsmath,amssymb,amsfonts}%
\usepackage{amsthm}%
\usepackage{mathrsfs}%
\usepackage[title]{appendix}%
\usepackage{xcolor}%
\usepackage{textcomp}%
\usepackage{manyfoot}%
\usepackage{booktabs}%
\usepackage{algorithm}%
\usepackage{algorithmicx}%
\usepackage{algpseudocode}%
\usepackage{listings}%

\begin{document}

\title{Potential Impact of the Sagittarius Dwarf Galaxy on the Formation of Young O-rich Stars}






\author[1,2,3]{\fnm{Tiancheng} \sur{Sun}}

\author*[1,2]{\fnm{Shaolan} \sur{Bi}}\email{bisl@bnu.edu.cn}

\author*[4,5]{\fnm{Xunzhou} \sur{Chen}}\email{cxzchen123@gmail.com}

\author[3,1,6]{\fnm{Yuqin} \sur{Chen}}

\author[7,8,9]{\fnm{Yuxi (Lucy)} \sur{Lu}}

\author[3,1,6]{\fnm{Chao} \sur{Liu}}

\author[10,11]{\fnm{Tobias} \sur{Buck}}

\author[1,2]{\fnm{Xianfei} \sur{Zhang}}

\author[1,2]{\fnm{Tanda} \sur{Li}}

\author[12]{\fnm{Yaguang} \sur{Li}}

\author[3]{\fnm{Yaqian} \sur{Wu}}

\author[13]{\fnm{Zhishuai} \sur{Ge}}

\author[1,2]{\fnm{Lifei} \sur{Ye}}

\affil*[1]{\orgdiv{Institute for Frontiers in Astronomy and Astrophysics}, \orgname{Beijing Normal University}, \orgaddress{\city{Beijing}, \postcode{102206}, \country{China}}}

\affil[2]{\orgdiv{School of Physics and Astronomy}, \orgname{Beijing Normal University}, \orgaddress{\city{Beijing}, \postcode{100875}, \country{China}}}

\affil[3]{\orgdiv{CAS Key Laboratory of Optical Astronomy}, \orgname{National Astronomical Observatories}, \orgaddress{\city{Beijing}, \postcode{100101}, \country{China}}}

\affil[4]{\orgdiv{Research Center for Astronomical Computing}, \orgname{Zhejiang Laboratory}, \orgaddress{\city{Hangzhou}, \postcode{311100}, \country{China}}}

\affil[5]{\orgdiv{School of Science}, \orgname{Hangzhou Dianzi University}, \orgaddress{\city{Hangzhou}, \postcode{310018}, \country{P. R. China}}}

\affil[6]{\orgname{School of Astronomy and Space Science, University of Chinese Academy of Sciences}, \orgaddress{\city{Beijing}, \postcode{100049}, \country{China}}}

\affil[7]{\orgname{American Museum of Natural History}, \orgaddress{\city{Central Park West, Manhattan, NY}, \postcode{10024}, \country{USA}}}

\affil[8]{\orgdiv{Department of Astronomy}, \orgname{The Ohio State University}, \orgaddress{\city{140 W 18th Ave, Columbus, OH}, \postcode{43210}, \country{USA}}}

\affil[9]{\orgdiv{Center for Cosmology and Astroparticle Physics (CCAPP)}, \orgname{The Ohio State University}, \orgaddress{\city{191 W. Woodruff Ave., Columbus, OH}, \postcode{43210}, \country{USA}}}

\affil[10]{\orgdiv{Universit\"at Heidelberg}, \orgname{ Interdisziplin\"ares Zentrum f\"ur Wissenschaftliches Rechnen}, \orgaddress{\city{Im Neuenheimer Feld 205, Heidelberg}, \postcode{D-69120}, \country{Germany}}}

\affil[11]{\orgdiv{Universit\"at Heidelberg}, \orgname{Zentrum f\"ur Astronomie, Institut f\"ur Theoretische Astrophysik}, \orgaddress{\city{Albert-Ueberle-Stra$\beta$e 2, Heidelberg}, \postcode{D-69120}, \country{Germany}}}

\affil[12]{\orgdiv{Institute for Astronomy}, \orgname{University of Hawai'i}, \orgaddress{\city{2680 Woodlawn Drive, Honolulu, HI}, \postcode{96822}, \country{USA}}}

\affil[13]{\orgdiv{Beijing Planetarium}, \orgname{Beijing Academy of Science and Technology}, \orgaddress{\city{Beijing}, \postcode{100044}, \country{China}}}


\abstract{The Milky Way underwent significant transformations in its early history, characterised by violent mergers and satellite galaxy accretion. However, recent observations reveal notable star formation events over the past 4 Gyr, likely triggered by perturbations from the Sagittarius dwarf galaxy. Here, we present chemical signatures of this accretion event, using the [Fe/H] (metallicity) and [O/Fe] (oxygen abundance) ratios of thin-disc stars. In the normalised age-metallicity plane, we identify a discontinuous V-shape structure at z$_{\rm max}$ (maximum vertical distance from the disc plane) $<$ 0.4 kpc in the local disc, interrupted by a star formation burst between 4 and 2 Gyr ago. This event is characterised by a significant increase in oxygen abundance, resulting in a distinct [O/Fe] gradient and the formation of young O-rich stars. These stars have larger birth radii, indicating formation in the outer disc followed by radial migration to the Solar neighbourhood. Simulations of late satellite infall suggest that the passage of the Sagittarius dwarf galaxy may have contributed to the observed increase in oxygen abundance in the local disc.}




\maketitle

\section*{Introduction}\label{sec1}

\begin{figure*}[ht]
\includegraphics[scale=1]{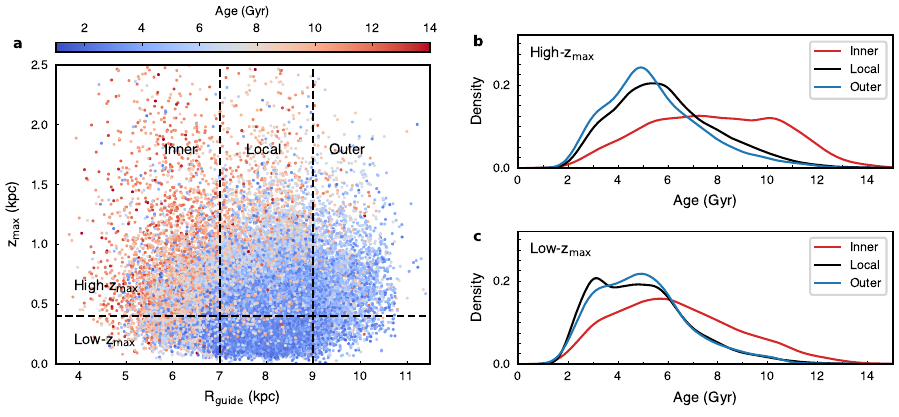}
\caption{\textbf{The age distributions of our star sample. a}, R$_{\rm guide}$ (guiding radius)-z$_{\rm max}$ (maximum vertical distance from the disc plane) distributions of sample stars, colour-coded by stellar ages. The vertical dashed lines in the top panel indicate the division into three R$_{\rm guide}$ bins (inner, local, and outer) at R$_{\rm guide}$ = 7 and 9 kpc. The horizontal dashed line indicates the division of each R$_{\rm guide}$ bin into two z$_{\rm max}$ bins (high-z$_{\rm max}$ and low-z$_{\rm max}$) at z$_{\rm max}$ = 0.4 kpc. \textbf{b}, The age distributions (density distributions, based on the kernel density estimates) of spatially selected subsamples (inner, local, and outer) in high-z$_{\rm max}$ region. \textbf{c}, The age distributions (density distributions) of spatially selected subsamples (inner, local, and outer) in low-z$_{\rm max}$ region. Source data are provided as a Source Data file.
\label{fig:rg_z}}
\end{figure*}

Recent studies using data from the European Space Agency (ESA) Gaia mission \cite{2018A&A...616A...1G,2023A&A...674A...1G} and the Galactic Archaeology with HERMES (GALAH) \cite{2015MNRAS.449.2604D,2018MNRAS.478.4513B,2021MNRAS.506..150B} survey have identified an enhanced star formation rate in the Galactic disc over the past 2-4 Gyr \cite{2019A&A...624L...1M,2019ApJ...878L..11I,2022MNRAS.510.4669S,2023MNRAS.523.1199S}, corroborating earlier findings from Hipparcos data  \cite{2000MNRAS.316..605H}. This star formation burst is likely linked to the pericentric passages of the Sagittarius dwarf galaxy (Sgr) \cite{2018MNRAS.478.5263T,2020NatAs...4..965R,2022arXiv221017054A}, which have potentially influenced the radial abundance gradient \cite{2023MNRAS.tmp.1561R} and the age-metallicity relation in the Milky Way disc \cite{2022MNRAS.512.4697L}.


Main-sequence turn-off (MSTO) and subgiant stars serve as reliable tracers of Galactic populations \cite{2017RAA....17....5W,2022ApJ...929..124C,2022Natur.603..599X}. With precise parallaxes from Gaia DR3 \cite{2023A&A...674A...1G} , the ages of these stars can be determined with an uncertainty of less than 10 percent  \cite{2023MNRAS.523.1199S}. 
Here we investigate the impact of recent accretion events on the star formation history of the Milky Way disc, using an oxygen-enhanced stellar model \cite{2023MNRAS.523.1199S} to determine the ages of MSTO and subgiant stars from the Third Data Release of GALAH \cite{2021MNRAS.506..150B} (GALAH DR3). Our approach utilises a Bayesian methodology \cite{2010ApJ...710.1596B}, incorporating spectroscopic chemical abundances, specifically metallicity ([Fe/H]), $\alpha$ abundance ([$\alpha$/Fe], where $\alpha$ refers to Mg, Si, Ca, and Ti), and oxygen abundance ([O/Fe]), alongside effective temperature T$_{\rm eff}$ and luminosity \cite{2023ApJS..264...41Y}.
The application of the oxygen-enhanced stellar model together with precise luminosity from Gaia and reliable abundance measurements from GALAH enable us to ascertain the age-abundance relations and track the evolution of the abundance gradient in the Galactic disc with high precision.

\section*{Results}\label{sec20}
\subsection*{Age–abundance distribution of the Milky Way disc}\label{sec2}

\begin{figure*}[ht]
\includegraphics[scale=1]{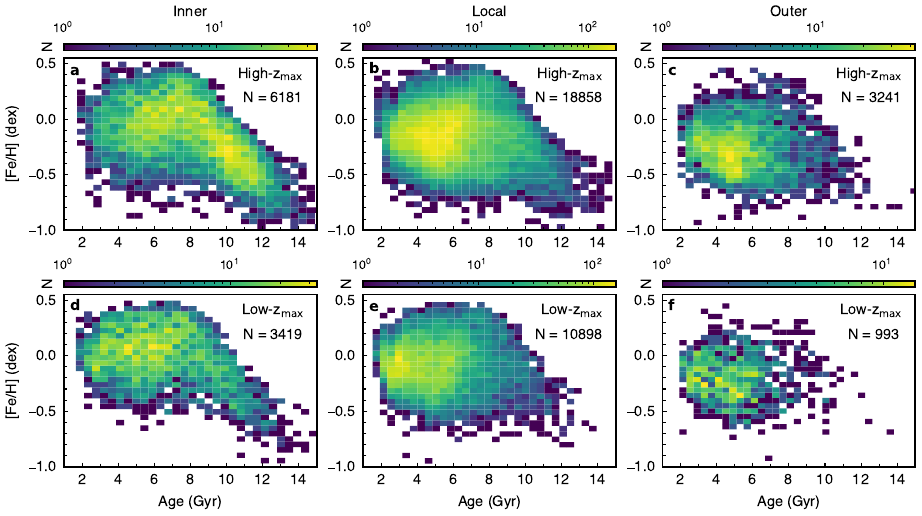}
\caption{\textbf{Age-[Fe/H] distributions of the six spatially selected subsamples.} 
\textbf{a-f}, arranged according to the division in Fig. \ref{fig:rg_z}a, colour-coded by the stellar number density, N. The first row of the figure corresponds to the high-z$_{\rm max}$ region, while the second row represents the low-z$_{\rm max}$ region. From left to right, the columns depict the inner, local, and outer regions, respectively. As shown in Fig. \ref{fig:rg_z}a, high-z$_{\rm max}$ refers to z$_{\rm max}$ $>$ 0.4 kpc, while low-z$_{\rm max}$ indicates z$_{\rm max}$ $<$ 0.4 kpc. Inner represents R$_{\rm guide}$ $<$ 7 kpc, local corresponds to 7 kpc $<$ R$_{\rm guide}$ $<$ 9 kpc, and outer refers to R$_{\rm guide}$ $>$ 9 kpc. The numbers of stars in each bin are shown in the top-right corner of each panel. Source data are provided as a Source Data file. 
\label{fig:rz_sub_12fig}}
\end{figure*}

\begin{figure*}[ht!]
\includegraphics[scale=1]{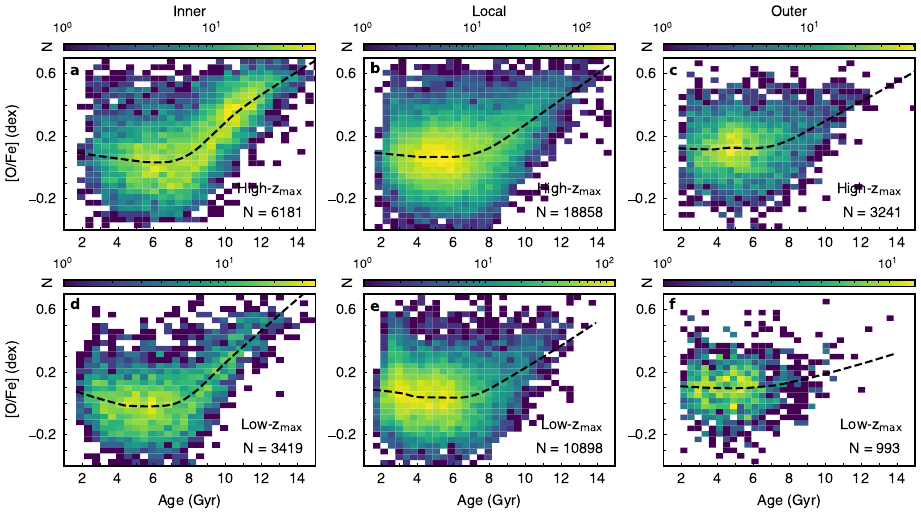}
\caption{\textbf{Age-[O/Fe] distributions of the six spatially selected subsamples.} \textbf{a-f}, arranged according to the division in Fig. \ref{fig:rg_z}, colour-coded by the stellar number density, N. The first row of the figure corresponds to the high-z$_{\rm max}$ region, while the second row represents the low-z$_{\rm max}$ region. From left to right, the columns depict the inner, local, and outer regions, respectively. As shown in Fig. \ref{fig:rg_z}a, high-z$_{\rm max}$ refers to z$_{\rm max}$ $>$ 0.4 kpc, while low-z$_{\rm max}$ indicates z$_{\rm max}$ $<$ 0.4 kpc. Inner represents R$_{\rm guide}$ $<$ 7 kpc, local corresponds to 7 kpc $<$ R$_{\rm guide}$ $<$ 9 kpc, and outer refers to R$_{\rm guide}$ $>$ 9 kpc. The numbers of stars in each bin are shown in the bottom-right corner of each panel. The black dashed lines represent the fitting result by local nonparametric regression. Source data are provided as a Source Data file.
\label{fig:rz_sub_12fig_o_fe}}
\end{figure*}

\begin{figure*}[ht!]
\includegraphics[scale=1]{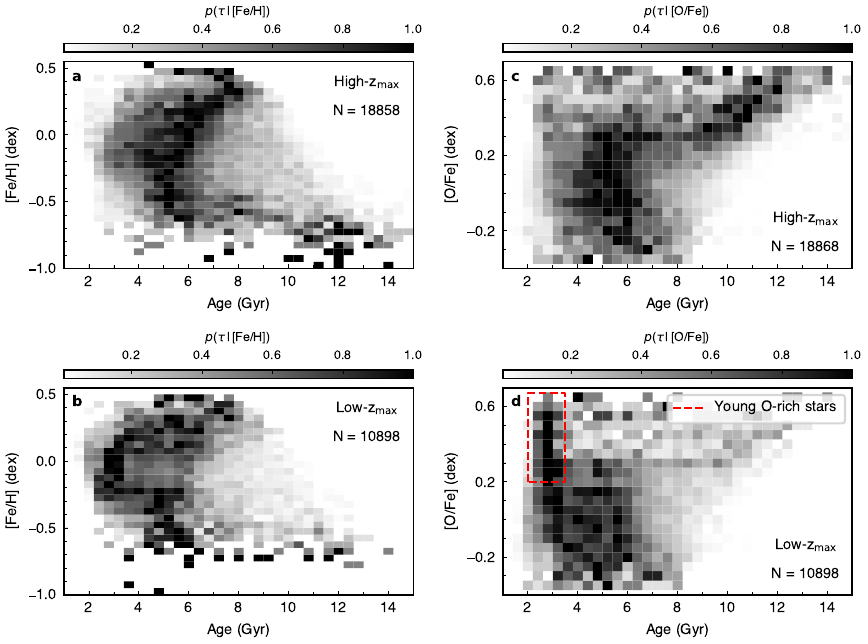}
\caption{\textbf{Stellar age–abundance relation of local disc revealed by our star sample.} \textbf{a}, \textbf{b}, Age-[Fe/H] distributions of the local disc stars at high-z$_{\rm max}$ and low-z$_{\rm max}$ regions, according to the division in Fig.\ref{fig:rg_z}a. 
\textbf{c}, \textbf{d}, Age-[O/Fe] distributions of the local disc stars at high-z$_{\rm max}$ and low-z$_{\rm max}$ regions. The red dashed box indicates the young O-rich stars.
\textbf{a}, Probability distribution of stellar age p($\tau$$\mid$[Fe/H]), normalised to the peak value for each [Fe/H], for local disc stars at high-z$_{\rm max}$ region. 
\textbf{b}, Similar to \textbf{a} but for local disc stars at low-z$_{\rm max}$ region.
\textbf{c}, Probability distribution of stellar age p($\tau$$\mid$[O/Fe]), normalised to the peak value for each [O/Fe], for local disc stars at high-z$_{\rm max}$ region. 
\textbf{d}, Similar to \textbf{c} but for local disc stars at low-z$_{\rm max}$ region. Source data are provided as a Source Data file.
\label{fig:rg_feh}}
\end{figure*}

Fig.\ref{fig:rg_z} shows the R$_{\rm guide}$ (guiding radius) versus z$_{\rm max}$ (maximum vertical distance from the disc plane) diagram and the derived age distributions of spatially selected subsamples. There are clear differences in age distributions (Fig.\ref{fig:rg_z}b,c) with R$_{\rm guide}$ and z$_{\rm max}$.
The ages of disc components (inner, local, and outer disc) at low-z$_{\rm max}$ region are generally younger than those at high-z$_{\rm max}$ regions, with the majority of stars being younger than 8 Gyr.
Fig.\ref{fig:rg_z}c shows a young peak at approximately 3 Gyr in the age distributions of local and outer disc at low-z$_{\rm max}$ region, which is more prominent for local disc. This bump nearly disappears in the high-z$_{\rm max}$ region (Fig.\ref{fig:rg_z}b). In addition, most of subsamples have an age peak at 5-6 Gyr, except for the inner disc at high-z$_{\rm max}$ region, which has a flat age distribution.
The young peak of age distributions indicate that there is a recent burst of star formation in the local and outer disc 3 Gyr ago, while the intermediate-aged peak at 5-6 Gyr is thought to be the star formation triggered by the first pericentric passages of Sgr (about 5.5 Gyr ago) \cite{2020NatAs...4..965R}.

The distributions of six spatially selected subsamples in age-[Fe/H] and age-[O/Fe] planes are presented in Fig.\ref{fig:rz_sub_12fig} and Fig.\ref{fig:rz_sub_12fig_o_fe}. Fig.\ref{fig:rz_sub_12fig_o_fe}a,b,d,e shows that there is an increasing trend of [O/Fe] with decreasing age in young (age $<$ 3 Gyr) and intermediate (4 Gyr $<$age $<$ 6 Gyr) populations. The oxygen-enhancement of young populations (hereafter young O-rich stars) is most prominent in local age-[O/Fe] relations (Fig.\ref{fig:rz_sub_12fig_o_fe}e), which correspond to the overdensities in local age-[Fe/H] relations (Fig.\ref{fig:rz_sub_12fig}e).
This result is consistent with the prediction of the late-burst and outer-burst galactic chemical evolution models from previous work \cite{2021MNRAS.508.4484J}, which suggest a approximately 0.1 dex uptick of [O/Fe] at ages of 2 Gyr and a V-shape form of age-metallicity relation. According to their models, there is an enhanced star formation rate 2 Gyr ago.

To better investigate the chemical enrichment history of the local disc, we employ a normalization procedure for the distribution $p$($\tau$,[Fe/H]) (the distribution of stellar age $\tau$ and metallicity [Fe/H]) of local disc stars to obtain $p$($\tau$$\mid$[Fe/H]) \cite{2022Natur.603..599X}, the age distribution at a specified [Fe/H]. 
As shown in Fig.\ref{fig:rg_feh}a,b, the resulting $p$($\tau$$\mid$[Fe/H]) distribution of local disc exhibit a gradual variation with z$_{\rm max}$, from a so called V-shape \cite{2018MNRAS.477.2326F,2019MNRAS.489.1742F,2021MNRAS.508.4484J,2022MNRAS.512.2890L,2022MNRAS.512.4697L,2022Natur.603..599X,2023MNRAS.523.1199S} (at age $<$ 8 Gyr) at high-z$_{\rm max}$ to a less discontinuous structure at low-z$_{\rm max}$ region.
This feature implies that the previously identified V-shape structure\cite{2022Natur.603..599X,2023MNRAS.523.1199S} depends on the location in the Milky Way disc, 
confirming early results \cite{2019MNRAS.489.1742F,2022MNRAS.512.2890L} based on APOGEE (Apache Point Observatory Galactic Evolution Experiment)\cite{2017AJ....154...94M} red giant star sample. For the local disc at high-z$_{\rm max}$ region, the distribution of $p$($\tau$$\mid$[Fe/H]) exhibits a V-shape. In the low-z$_{\rm max}$ region, the V-shape structure becomes discontinuous, and this discontinuity corresponds to a decrease (Fig.\ref{fig:rg_feh}b) in iron abundance and an sharp increase (Fig.\ref{fig:rg_feh}d) in oxygen abundance, suggesting that a fresh gas interrupted the secular evolution of the Milky Way disc \cite{2023A&A...670A.109S}.

\begin{figure*}[ht!]
\includegraphics[scale=1]{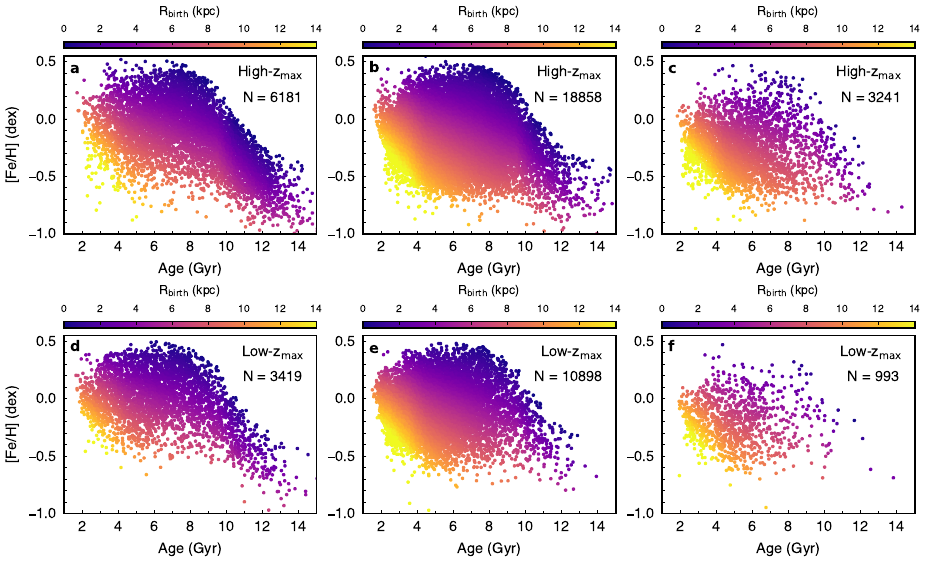}
\caption{\textbf{Age-[Fe/H] distributions of the six spatially selected subsamples, colour-coded by birth radius R$_{\rm birth}$.} 
\textbf{a-f}, arranged according to the division in Fig. \ref{fig:rg_z}a. The first row of the figure corresponds to the high-z$_{\rm max}$ region, while the second row represents the low-z$_{\rm max}$ region. From left to right, the columns depict the inner, local, and outer regions, respectively. As shown in Fig. \ref{fig:rg_z}a, high-z$_{\rm max}$ refers to z$_{\rm max}$ $>$ 0.4 kpc, while low-z$_{\rm max}$ indicates z$_{\rm max}$ $<$ 0.4 kpc. Inner represents R$_{\rm guide}$ $<$ 7 kpc, local corresponds to 7 kpc $<$ R$_{\rm guide}$ $<$ 9 kpc, and outer refers to R$_{\rm guide}$ $>$ 9 kpc. The numbers of stars in each bin are shown in the top-right corner of each panel. Source data are provided as a Source Data file.
\label{fig:rz_sub_12fig_rb}}
\end{figure*}

\begin{figure*}[ht!]
\includegraphics[scale=1]{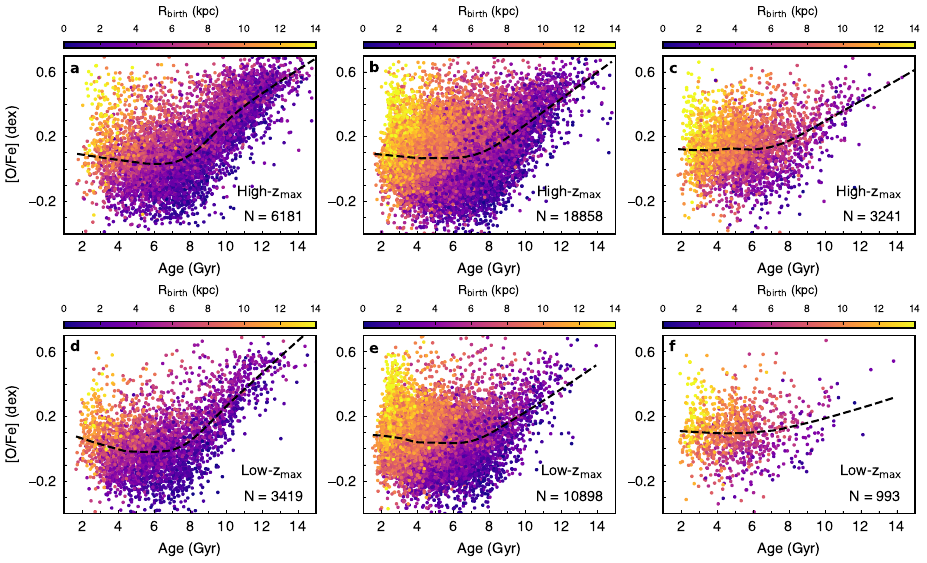}
\caption{\textbf{Age-[O/Fe] distributions of the six spatially selected subsamples, colour-coded by birth radius R$_{\rm birth}$.} \textbf{a-f}, arranged according to the division in Fig. \ref{fig:rg_z}. The first row of the figure corresponds to the high-z$_{\rm max}$ region, while the second row represents the low-z$_{\rm max}$ region. From left to right, the columns depict the inner, local, and outer regions, respectively. As shown in Fig. \ref{fig:rg_z}a, high-z$_{\rm max}$ refers to z$_{\rm max}$ $>$ 0.4 kpc, while low-z$_{\rm max}$ indicates z$_{\rm max}$ $<$ 0.4 kpc. Inner represents R$_{\rm guide}$ $<$ 7 kpc, local corresponds to 7 kpc $<$ R$_{\rm guide}$ $<$ 9 kpc, and outer refers to R$_{\rm guide}$ $>$ 9 kpc. The numbers of stars in each bin are shown in the bottom-right corner of each panel.
The black dashed lines represent the fitting result by local nonparametric regression. Source data are provided as a Source Data file.
\label{fig:rz_sub_12fig_o_fe_rb}}
\end{figure*}

Some of previous studies \cite{2018MNRAS.477.2326F,2019MNRAS.489.1742F,2022Natur.603..599X} have suggested that V-shape structure is mainly caused by radial migration, i.e., the metal-rich stars likely migrated from the inner disk to the Solar neighbourhood, while the metal-poor branch arises from stars that were born in the outer disc and have migrated inwards. However, other works \cite{2021MNRAS.508.4484J,2022MNRAS.512.2890L,2022MNRAS.512.4697L} based on both the observations and simulations indicated that the late satellite infall as the cause of the V shape. In this scenario, the late satellite infall could triggers star formation burst in the outer disc, and the radial migration subsequently brings the stars born from the infallen gas into the local disc. 
A direct approach for studying the effect of radial migration on the V-shape structure is to infer the birth radii \cite{2018MNRAS.481.1645M,2022arXiv221204515L}(see methods, subsection Calculation of Birth Radius) of the stars, and the results are presented in Fig.\ref{fig:rz_sub_12fig_rb} and Fig.\ref{fig:rz_sub_12fig_o_fe_rb}. 

\begin{figure*}[ht!]
\includegraphics[scale=1]{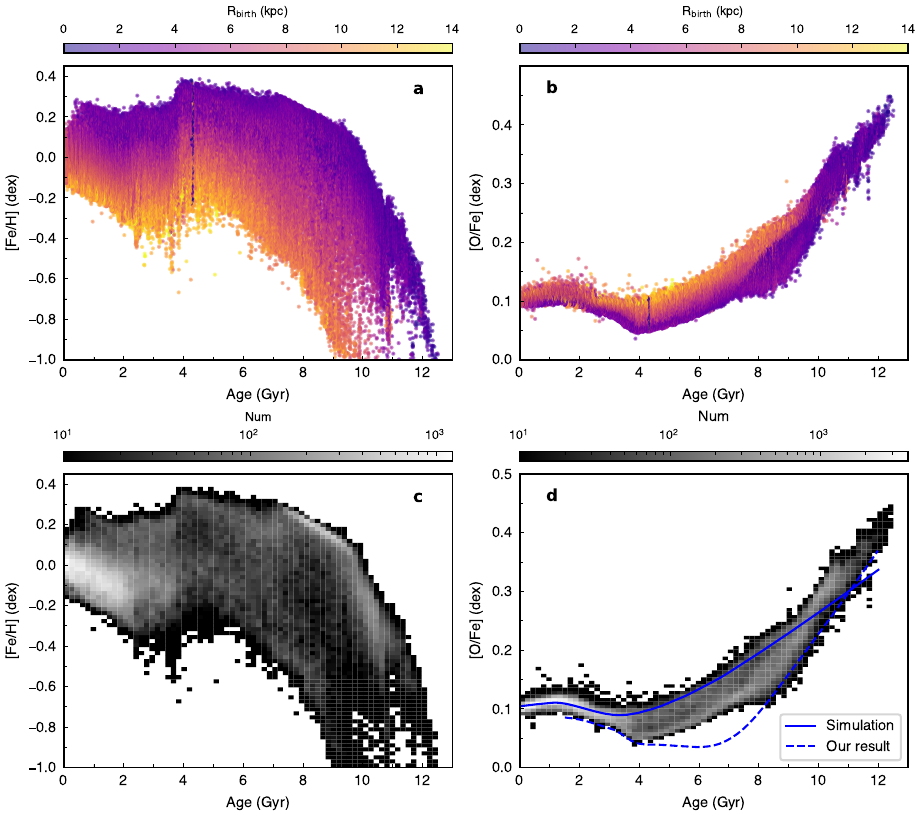}
\caption{\textbf{The effect of radial migration after the satellite infall on the age-[Fe/H]/[O/Fe] in the simulation, restricted to the solar neighbourhood (vertical distance from the disc plane $|$Z$_{\rm Gal}$$|$ $<$ 0.4 kpc and 7 kpc $<$ R$_{\rm Gal}$ $<$ 9 kpc, where R$_{\rm Gal}$ represents the Galactocentric radius).} 
\textbf{a-b}, colour-coded by birth radius R$_{\rm birth}$. \textbf{c-d}, colour-coded by the stellar number density. The blue solid lines represent the fitting result by local nonparametric regression, and the blud dashed lines (the fitting results for local disc at low-z$_{\rm max}$ region, shown in Fig.\ref{fig:rz_sub_12fig_rb}e and Fig.\ref{fig:rz_sub_12fig_o_fe_rb}e) are overplotted for comparison. Source data are provided as a Source Data file.
\label{fig:simulation}}
\end{figure*}

Fig.\ref{fig:rz_sub_12fig_rb} shows that the sub-solar metallicity stars with age $<$ 6 Gyr have R$_{\rm birth}$ (birth radius) large than 12 kpc, suggesting a outer disc formation and subsequent radial migration to the Solar neighbourhood. In the same epoch, the birth radii of metal-rich stars are less than 6 kpc, indicating that they are born in the inner disc. Intriguingly, the most of young O-rich stars found in Fig.\ref{fig:rz_sub_12fig_o_fe} have R$_{\rm birth}$ large than 12 kpc (Fig.\ref{fig:rz_sub_12fig_o_fe_rb}), which means the stars migrated from the outer disc are oxygen-enhanced as well.
This result seems to be consistent with the outer-burst galactic chemical evolution model \cite{2021MNRAS.508.4484J}, which assumes that local and outer disc experience a starburst 2 Gyr ago, with inner regions following the inside–out star formation histories. Additionally, Lu et al. (2022) \cite{2022MNRAS.512.4697L} examined the impact of a Sagittarius-like galaxy passage on the V-shaped structure of Milky Way analogues. We compare their simulated age-[Fe/H]/[O/Fe] distributions in the Solar neighbourhood (vertical distance from the disc
plane $|$Z$_{\rm Gal}$$|$ $<$ 0.4 kpc and 7 kpc $<$ R$_{\rm Gal}$ $<$ 9 kpc, where R$_{\rm Gal}$ represents the Galactocentric radius) with our results.
As shown in Fig.\ref{fig:simulation}a,b, according to the simulations, the lower envelope of age-[Fe/H] distribution and the upper envelope of age-[O/Fe] distribution arise from the stars born in the outer disc, which is similar with our results in Fig.\ref{fig:rz_sub_12fig_rb} and Fig.\ref{fig:rz_sub_12fig_o_fe_rb}. 
The simulations reveal a notable discontinuity in the age-[Fe/H] relation at 4 Gyr, attributable to the influence of a Sagittarius-like galaxy merger event \cite{2022MNRAS.512.4697L}. This event reduces the metallicity of the interstellar medium (ISM), causing stars formed after 4 Gyr ago to inherit lower metallicity than older stars.
Fig.\ref{fig:simulation}d shows that the increasing trend of [O/Fe] with younger age at age $<$ 4 Gyr is in reasonable agreement with our results. Compared with the simulations, our sample focuses on the more metal-rich disc stars, resulting in an overall lower oxygen abundance than the simulations.
The comparison between our results and the simulations \cite{2022MNRAS.512.4697L} suggests that the passage of the Sagittarius dwarf galaxy may have contributed to an elevated star formation rate in the outer disc. This increase in star formation rate likely led to the emergence of newly formed O-rich stars in this region. Subsequently, radial migration guiding these newly formed stars to the Solar neighbourhood.

\begin{figure*}[ht!]
\includegraphics[scale=1]{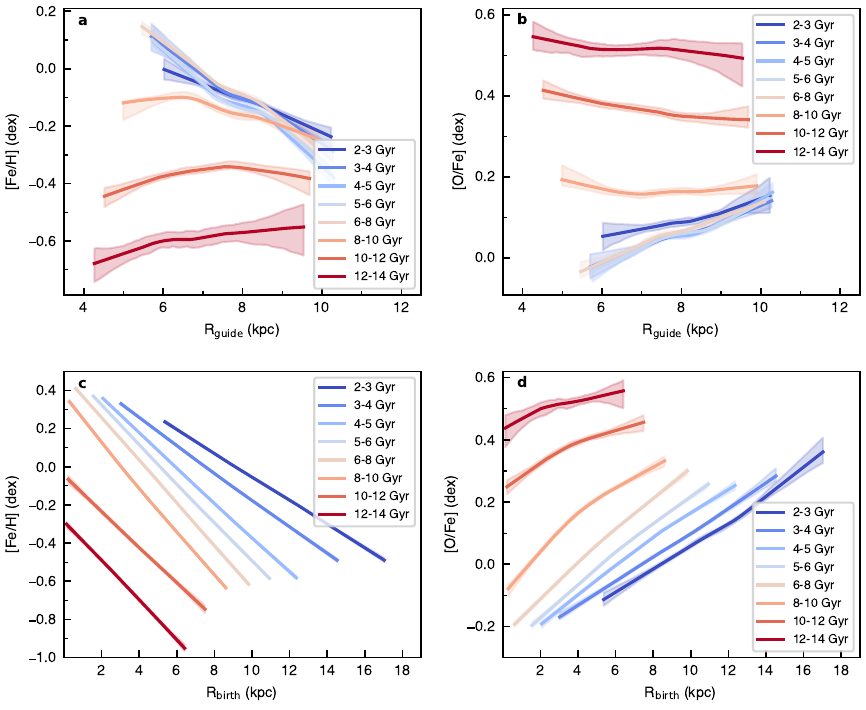}
\caption{\textbf{Radial abundance profile in bins of age.} \textbf{a}, Radial [Fe/H] profile in bins of age, with respect to guiding radius R$_{\rm guide}$, each line represent the local nonparametric regression fitting to the distribution of sample stars in this age bins. The shaded regions indicate the 95\% confidence interval around the fitting result by performing bootstrap resampling. \textbf{b}, Similar to \textbf{a} but for radial [O/Fe] profile in bins of age.
\textbf{c}, Radial [Fe/H] profile in bins of age, with respect to birth radius R$_{\rm birth}$. \textbf{d}, Similar to \textbf{c} but for radial [O/Fe] profile. Source data are provided as a Source Data file.
\label{fig:rg_feh_p}}
\end{figure*}

\subsection*{Temporal evolution of radial abundance gradient}\label{sec3}

\begin{figure*}[ht!]
\includegraphics[scale=1]{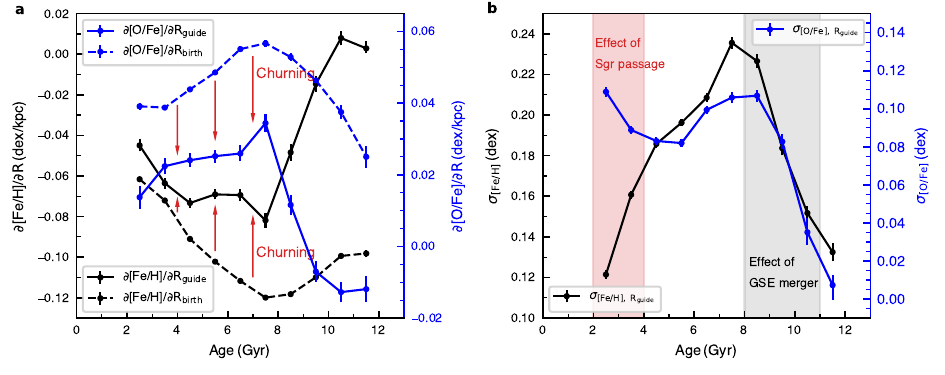}
\caption{\textbf{Age dependence of the radial abundance gradient ($\partial$[Fe/H]/$\partial {\rm R}$) and the corresponding abundance dispersion ($\sigma_{\rm [Fe/H]}$) around the gradient.} 
The uncertainties (error bars) in each panel are derived using MCMC (Markov Chain Monte Carlo) fitting, employing the Python code emcee \cite{2013ascl.soft03002F}.
\textbf{a}, Age dependence of the radial [Fe/H] ($\partial$[Fe/H]/$\partial \rm R_{\rm guide}$, black solid line) gradient and radial [O/Fe] ($\partial$[O/Fe]/$\partial \rm R_{\rm guide}$, blue solid line) gradient, in terms of guiding-centre radius R$_{\rm guide}$. Each point was obtained by 3-parameter (slope, intercept, and dispersion) Bayesian fits to the [Fe/H]/[O/Fe]-R$_{\rm guide}$ distribution, using only data in the respective age bin. The black and blue dashed lines represent the radial [Fe/H]/[O/Fe] gradients ($\partial$[Fe/H]/$\partial \rm R_{\rm birth}$ and $\partial$[O/Fe]/$\partial \rm R_{\rm birth}$) in terms of birth radius R$_{\rm birth}$, calculated for the same age bins. The difference between the dashed lines and the solid lines can be attributed the influence of radial migration (churning). 
\textbf{b}, Age dependence of the [Fe/H] ($\sigma_{{\rm [Fe/H]},\ \rm R_{\rm guide}}$, black) and [O/Fe] ($\sigma_{{\rm [O/Fe]},\ \rm R_{\rm guide}}$, blue) dispersion around the radial [Fe/H]/[O/Fe] gradient (\textbf{a}), in terms of R$_{\rm guide}$. The grey-shaded area marks the age interval in which we expect to see signatures from the Gaia Sausage/Enceladus (GSE) merger event, while the red-shaded area marks the age interval for the effect of Sagittarius dwarf galaxy (Sgr) passage. Source data are provided as a Source Data file. 
\label{fig:rg_feh_delta_fe}}
\end{figure*}

\begin{figure*}[ht!]
\includegraphics[scale=1]{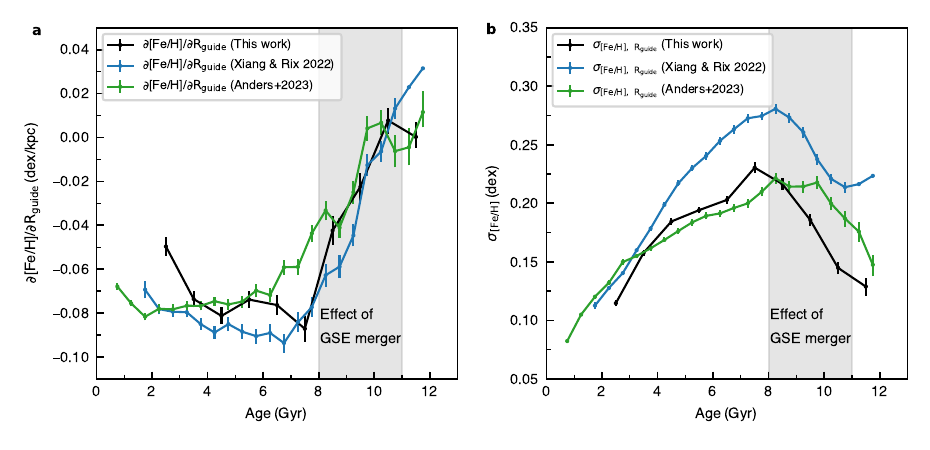}
\caption{\textbf{Comparison of the radial [Fe/H] gradient ($\partial$[Fe/H]/$\partial \rm R_{\rm guide}$, in terms of guiding-centre radius R$_{\rm guide}$) and the corresponding [Fe/H] dispersion ($\sigma_{\rm [Fe/H]}$) around the gradient in this work, with the results from LAMOST and APOGEE.} 
The uncertainties (error bars) in each panel are derived using MCMC (Markov Chain Monte Carlo) fitting, employing the Python code emcee \cite{2013ascl.soft03002F}.
\textbf{a}, \textbf{b}, The black lines in each panel represent the result of this work, using the GALAH subgiant and MSTO stars; the blue and green lines represent the results from the LAMOST data (LAMOST DR7 subgiant sample from Xiang \& Rix 2022 \cite{2022Natur.603..599X}) and literature (based on APOGEE DR17 red giant sample from Anders et al. 2023 \cite{2023arXiv230408276A}), respectively. The grey-shaded area marks the age interval in which we expect to see signatures from the Gaia Sausage/Enceladus (GSE) merger event. Source data are provided as a Source Data file. 
\label{fig:rz_o_evo_radial}}
\end{figure*}

Previous studies on the Gaia-Enceladus/Sausage (GSE) have highlighted the significant role of massive mergers in shaping the Galactic disc and altering the radial metallicity gradient in the past 8-11 Gyr ago \cite{2021SCPMA..6439562Z,2022arXiv221204515L,2023arXiv230408276A,2023MNRAS.tmp.1561R}.
To study the influence of potential minor merger events (Sgr accretion) on the radial metallicity gradient during the later stages of Galaxy evolution. 
We examine the temporal evolution of the radial abundance gradient in the Milky Way disc by utilising precise stellar ages from our stellar models.

Fig.\ref{fig:rg_feh_p}a,b presents the radial profiles of [Fe/H] and [O/Fe] for disc stars with respect to the guiding-center radius R$_{\rm guide}$, divided into eight age bins. Notably, the [Fe/H] profile (Fig.\ref{fig:rg_feh_p}a) undergoes a substantial transformation, transitioning from a positive gradient at 12-14 Gyr to a negative gradient at 6-8 Gyr, and subsequently maintaining a relatively steady negative gradient until 3 Gyr ago. A significant departure from this stable gradient occurs at 2-3 Gyr, as the [Fe/H] profile becomes flatter compared to the 3-8 Gyr period and resembles the gradient observed at 8-10 Gyr.

The radial [Fe/H] profile at 10-12 Gyr exhibits a break at about 7.5 kpc, featuring a positive slope within the break radius and a negative slope beyond it. Similarly, the radial [Fe/H] profile at 8-10 Gyr ago shows a break at approximately 6.5 kpc, with a flat slope within the break radius and a negative slope beyond it. 
These break radii is consistent with the break radius observed in the radial profile of integrated stellar metallicity using red giant branch stars \cite{2023arXiv230614100L}. 
Moreover, leveraging the high precision in age, we identify that these breaks primarily occur within the 8-12 Gyr, which was not apparent in the metallicity profiles of mono-age populations in previous studies \cite{2023arXiv230614100L}.

The radial [O/Fe] profile (Fig.\ref{fig:rg_feh_p}b) exhibits a transition from a negative gradient at 12-14 Gyr to a positive gradient at 6-8 Gyr, followed by a relatively stable positive gradient until 3 Gyr ago. Similar to the radial [Fe/H] profile, a significant departure from this stable radial [O/Fe] profile occurs at 2-3 Gyr, with a flatter gradient compared to the 3-8 Gyr period. Overall, the flattened radial [Fe/H] and radial [O/Fe] profiles imply that an accretion event has diluted the metallicity of the disc and led to an enhancement of oxygen abundance.

Since the guiding radius (R$_{\rm guide}$) is just an approximation to the birth radius of a star, it cannot remove the effect of churning that moves stars across the disc without significantly changing their orbital eccentricities \cite{2002MNRAS.336..785S}. We now investigate the radial abundance gradients in terms of birth radius R$_{\rm birth}$, to better record the time evolution of [Fe/H] and [O/Fe] in the Milky Way disc. 
In Fig.\ref{fig:rg_feh_p}c, the youngest population (2-3 Gyr) exhibits a slight narrow metallicity range (from about $-$0.5 dex to 0.2 dex) compared with older populations. This population also shows a clear signature of oxygen-enhancement in Fig.\ref{fig:rg_feh_p}d. As seen in the figure, from 14 Gyr ago to 4 Gyr ago, the oxygen abundance of mono-age populations decrease with lookback time. However, from 4 Gyr ago to 2 Gyr ago, there is a increase of oxygen abundance with lookback time, and the most oxygen-rich stars born in outer disc, supporting the results in Fig.\ref{fig:rz_sub_12fig_o_fe_rb}.

The Bayesian linear fitting \cite{2017A&A...600A..70A} are performed to the radial [Fe/H]/[O/Fe] profiles in 1 Gyr age bins (Fig.\ref{fig:rg_feh_delta_fe}) to present the distinctive characteristics of them. 
Fig.\ref{fig:rg_feh_delta_fe}a contrast the [Fe/H]/[O/Fe] gradient with R$_{\rm birth}$ to the gradient with R$_{\rm guide}$.
The present-day abundance gradients of mono-age populations in terms of R$_{\rm guide}$ are shown in black and blue solid lines, while the gradients in terms of R$_{\rm birth}$ are shown in black and blue dashed lines.
The difference between the solid and dashed curves results from radial migration (churning), which flattens the observed radial abundance gradient from 8 to 2 Gyr ago.
In Fig.\ref{fig:rg_feh_delta_fe}b, the enhanced dispersion in the radial [O/Fe] gradient after 4 Gyr, relative to the 6–4 Gyr period, suggests that an accretion event influenced the oxygen abundance of the thin disc via radial migration.
However, there is no obvious feature of enhanced dispersion about radial [Fe/H] gradient. 
The difference between the dispersion of [Fe/H] and [O/Fe] indicates that this accretion event introduced a small dispersion of [Fe/H] (the overdensity in Fig.\ref{fig:rz_sub_12fig}e). Furthermore, at the early stage (8-11 Gyr) of Milky Way, there is a sharp rise in the dispersion of [Fe/H]/[O/Fe] gradient, linked to the effect of the GSE merger event \cite{2022arXiv221204515L,2023MNRAS.tmp.1561R}, slighter later (by about 0.5 Gyr, see Fig.\ref{fig:rz_o_evo_radial}) than the epoch based on the LAMOST (Large Sky Area Multi-Object Fiber Spectroscopic Telescope) \cite{2012RAA....12.1197C,2012RAA....12..723Z} and APOGEE \cite{2017AJ....154...94M,2022ApJS..259...35A} data \cite{2022arXiv221204515L,2023arXiv230408276A}. These discrepancies could be attributed to different methods of age determination. 
Compared with the scaled-solar mixture stellar model, the $\alpha$-enhanced stellar model \cite{2000ApJ...532..430V,2001ApJ...556..322B,2001ApJS..136..417Y,2002ApJS..143..499K} yields significantly younger age estimates for thick disc stars. Moreover, Sun et al. (2023) \cite{2023MNRAS.523.1199S} found that using O-enhanced stellar model further reduces the age estimates of thick disc stars by approximately 5\% if [O/$\alpha$] = 0.2 dex.

In this work, we confirm that the V-shape age–[Fe/H] relation varies across different regions of the Galactic disc. This structure is disrupted at z$_{\rm max}$ $<$ 0.4 kpc, characterised by a decrease in metallicity ([Fe/H]) and a pronounced increase in oxygen abundance ([O/Fe]) between 4 and 2 Gyr ago. Based on the birth radii of the sample stars, we identify that these young O-rich stars predominantly originate from the outer disc and have migrated to the Solar neighbourhood through radial migration. These results are consistent with the simulations that attribute such features to the late infall of a satellite galaxy, likely resembling the Sagittarius dwarf, into the Milky Way. Additionally, we find that this satellite infall generates distinct radial [Fe/H] and [O/Fe] profiles compared with earlier epochs, along with a significant increase in the dispersion of the radial [O/Fe] gradient for younger stars formed during the same period (2–4 Gyr ago). These findings suggest that the infall of the Sagittarius dwarf galaxy can trigger star formation burst in the outer disc, contributing to the formation of young O-rich stars and reshaping the chemical composition of the Milky Way disc via radial migration.

\backmatter

\begin{appendices}

\section*{Methods}\label{secA1}

\subsection*{GALAH Data and Sample Selection}\label{data}

This work is based on the data from the Third Data Release of the Galactic Archaeology with HERMES survey (GALAH DR3) \cite{2021MNRAS.506..150B}.
GALAH DR3 \citep{2021MNRAS.506..150B} provides stellar parameters (effective temperature $T_{\rm eff}$, surface gravity $\log g$, metallicity [Fe/H], microturbulence velocity $V_{mic}$, broadening velocity $V_{broad}$, radial velocity $V_{rad}$) and up to 30 elemental abundances for 588,571 stars, derived from optical spectra at a typical resolution of R about 28,000. 
The metallicity [Fe/H], oxygen abundance [O/Fe], magnesium abundance [Mg/Fe], silicon abundance [Si/Fe], and calcium abundance [Ca/Fe] from GALAH DR3 was calculated based on a non-LTE method (LTE: local thermodynamic equilibrium) \cite{2020A&A...642A..62A}.
The data set used in this work is mainly from Sun et al. 2023 \cite{2023MNRAS.523.1199S}. We extended this sample \cite{2023MNRAS.523.1199S} to cover a $T_{\rm eff}$ range of 4800-6500 K, and a $\log g$ range of 3.2-4.1.
Following the recommendations in GALAH DR3, we apply stringent selection criteria to ensure reliable stellar parameters, including iron, $\alpha$-elements, and oxygen abundances (stellar parameter quality flag flag$\_$sp = 0, quality flag for the overall iron abundance flag$\_$fe$\_$h = 0, the $\alpha$ abundance flag$\_$alpha$\_$fe = 0, and the oxygen abundance flag$\_$o$\_$fe = 0), requiring an signal-to-noise ratio SNR $>$ 30, and a chi2$\_$sp $<$ 4 (Chi2 value of stellar parameter fitting). Binary systems identified by Traven et al. 2020 \cite{2020A&A...638A.145T} and Yu et al. 2023 \cite{2023ApJS..264...41Y} are excluded. Additionally, we apply a single cut based on the Gaia DR3 parameters by selecting stars with a Gaia re-normalised unit weight error (RUWE) of less than 1.2. Giant stars are excluded by applying the absolute magnitude cut \cite{2022MNRAS.510.4669S}:
\begin{equation}\label{e0}
\begin{split}
M_{K_{s}} =\,m_{K_{s}} - \,A_{K_{s}} - \,5\rm log10[(100\rm\,mas)/\varpi] >\\
8.5\,-\,T_{\rm eff}/(700\rm\,K)
\end{split}
\end{equation}
The extinction values $A_{K_{s}}$ and the 2MASS $m_{K_{s}}$ magnitudes \cite{2006AJ....131.1163S} used here are taken from the GALAH catalogue.
The parallax $\varpi$ come from GAIA DR3 catalogue \cite{2023A&A...674A..38G}.
In addition, we remove all stars with absolute magnitude $M_{\rm K}$ brighter than 0.5 mag to avoid contamination from He-burning horizontal branch stars \cite{2022Natur.603..599X}.
To focus on disc stars, we select samples with [Fe/H] $>$ $-$1, eccentricity $<$ 0.5, and $|$Z$_{\rm Gal}$$|$ $<$ 1 kpc, excluding the halo stars mentioned in Sun et al. 2023 \cite{2023MNRAS.523.1199S}. 
To ensure result accuracy, stars with relative age uncertainties exceeding 30 per cent were removed. Additionally, we exclude 2 stars with significant model systematic bias, whose inferred ages are 2 $\sigma$ larger than the age of the Universe (13.8 Gyr) \cite{2016A&A...594A..13P}. 
After applying these cuts, our final sample consisted of 43,590 MSTO and subgiant stars, with a median relative age uncertainty of 9.7 per cent across the age range of 1–13.8 Gyr, as shown in Supplementary Figure 1.
We obtain the luminosities of sample stars by cross-match them with the catalogue from Yu et al. 2023 \cite{2023ApJS..264...41Y}, which provides the luminosity of 1.5 million stars using astrometric data from GAIA DR3 \cite{2023A&A...674A..38G} and improved interstellar extinction measurements.

We utilised the orbital parameters (eccentricity) and velocities (Galactic rectangular x-velocity U, Galactic rectangular y-velocity V, Galactic rectangular z-velocity W, and Galactocentric vertical velocity V$_{\rm Z}$) from the GALAH DR3 value-added catalogue (VAC) \cite{2021MNRAS.506..150B}. These values are calculated from the astrometry provided by Gaia EDR3 and radial velocities determined from the GALAH spectra \cite{2021MNRAS.508.4202Z}. The orbital parameters in this catalogue are calculated using the Python package \texttt{Galpy} \cite{2015ApJS..216...29B}, with the assumed Milky Way potential and solar kinematic parameters detailed in Buder et al. 2021 \cite{2021MNRAS.506..150B}. We calculated the guiding radii R$_{\rm guide}$ with the same input parameters (distance, right ascension (ra), declination (dec), radial velocity, proper motion in right ascension direction (pmra), proper motion in declination direction (pmdec)), Milky Way potential, and solar kinematic parameters presented in Buder et al. 2021 \cite{2021MNRAS.506..150B}.

\subsection*{Age Estimation based on Oxygen-enhanced Stellar Models}\label{model}

The stellar ages used in this study are mainly from Sun et al. 2023 \cite{2023MNRAS.523.1199S}($T_{\rm eff}$ range of 5000-6500 K, and a $\log g$ range of 3.5-4.1). For stars with $T_{\rm eff}$ range of 4800-5000 K, or a $\log g$ range of 3.2-3.5, ages are estimated using the fitting method described in the Methods, subsection Fitting method.

We use oxygen-enhanced stellar evolution models to estimate ages of sample stars. 
The oxygen-enhanced stellar models use an individual O enhancement factor, thereby allowing the O abundance to be specified independently. The other $\alpha$-elements (i.e., Ne, Mg, Si, S, Ca, and Ti) are maintained with the same enhancement factor.
Neglecting to account for the independent enhancement of oxygen abundance in age determination would result in significant age biases, which would obscure the age-[O/Fe] relation \cite{2023ApJS..268...29S}. Therefore, the oxygen-enhanced models could accurately characterising the age-[O/Fe] relation of sample stars.

The ages of the MTSO and subgiant sample stars are determined by matching the Gaia Luminosity \cite{2023ApJS..264...41Y}, the GALAH spectroscopic stellar parameters $T_{\rm eff}$, [Fe/H], [$\alpha$/Fe], and [O/Fe], with the Oxygen-enhanced stellar models \cite{2023MNRAS.523.1199S} using a Bayesian approach \cite{2010ApJ...710.1596B}.

\subsubsection*{Fitting method}\label{Fitting}

In this study, we utilize five observed quantities, namely $T_{\rm eff}$, luminosity, [Fe/H], [$\alpha$/Fe], and [O/Fe], to determine stellar age. For each star, we use a set of stellar models with the corresponding [$\alpha$/Fe] and [O/Fe] (the models with [$\alpha$/Fe] and [O/Fe] values closest to the observed value of the star) to calculate stellar age. To find the most probable stellar models from evolutionary tracks, we adopt a 3-sigma error (i.e., three times the observational error) to define the likelihood that matches the observed constraints ($T_{\rm eff}$, [Fe/H], luminosity).

Following the fitting method introduced by \cite{2010ApJ...710.1596B}, we compare model predictions with their corresponding observational properties $D$ to calculate the overall probability of the model $M_i$ with posterior probability $I$,
\begin{equation}\label{e1}
p\left(M_{i}\mid D,I\right)=\frac{p\left(M_{i}\mid I\right) p\left(D\mid M_{i}, I\right)}{p(D\mid I)}
\end{equation}
where $p$($M_i$ $\mid$ $I$) represents the uniform prior probability for a specific model, and $p$(D $\mid$ $M_i$, $I$) is the likelihood function: 
\begin{equation}\label{e2}
\begin{aligned}
p\left(D\mid M_{i},I\right)=L(T_{\rm eff},\rm [Fe/H],\rm lum)\\
=L_{T_{\rm eff}}L_{\rm [Fe/H]}L_{\rm lum}
\end{aligned}
\end{equation}
L$_{T_{\rm eff}}$, L$_{\rm [Fe/H]}$, and L$_{\rm lum}$ represent the likelihood functions corresponding to the effective temperature $T_{\rm eff}$, metallicity [Fe/H], and luminosity, respectively.
The $p$($D$ $\mid$ $I$) in Eq. \ref{e1} is a normalization factor for the specific model probability:
\begin{equation}\label{e4}
p(D \mid I)=\sum_{j=1}^{N_{m}} p\left(M_{j} \mid I\right) p\left(D \mid M_{j}, I\right)
\end{equation}
where $N_m$ is the total number of selected models. The uniform priors $p$($M_i$ $\mid$ $I$) can be cancelled, giving the simplified Eq. \ref{e1} as :
\begin{equation}\label{e5}
p\left(M_{i} \mid D, I\right)=\frac{p\left(D \mid M_{i}, I\right)}{\sum_{j=1}^{N_{m}} p\left(D \mid M_{j}, I\right)}.
\end{equation}
We obtain the probability distribution for each star with Eq. \ref{e5} and fit a Gaussian function to the likelihood distribution.
The centre and standard deviation of the Gaussian profile are the estimate and uncertainty, respectively.

\subsubsection*{Age-Velocity Relation}\label{AVR}

Since our age estimates are independent from kinematics of sample stars. As an test for our age estimation, we show the age-velocity relation (AVR) of sample stars in Supplementary Figure 2. 
Supplementary Figure 2a shows the AVR of our sample in local region, with a Galactocentric distance between 7 kpc and 9 kpc.
Since the age range of our sample does not cover the youngest stars, we also plot the AVR recently obtained by Tarricq et al. 2021 \cite{2021A&A...647A..19T} using a sample of 418 Gaia-confirmed OCs (open clusters) in the solar neighbourhood. 
We note that the AVR at age $<$ 7 Gyr can be well described with a power law \cite{2016MNRAS.462.1697A,2018MNRAS.475.1093Y,2019ApJ...878...21T,2021MNRAS.503.5826A}, which is different from the power-law fitting of OC sample. 
Supplementary Figure 2b shows the AVR of local disc stars with a guiding radius between 7 kpc and 9 kpc. Compared with the result based on APOGEE DR17 red giants, our result is more consistent with the result from open clusters \cite{2021A&A...647A..19T}. Moreover, the AVR of our sample stars at age $<$ 7 Gyr is in good agreement with those of LAMOST subgiants \cite{2022Natur.603..599X}, indicating a good age precision of our sample stars.

\subsection*{Calculation of Birth Radius}\label{birth}


The birth radius R$_{\rm birth}$ is calculated following the method of Lu et al. 2022 \cite{2022arXiv221204515L}, an enhancement of the approach introduced by Minchev et al. 2018 \cite{2018MNRAS.481.1645M}. Lu et al. 2022 \cite{2022arXiv221204515L} established a linear relation between the ISM metallicity gradient at a given lookback time (\mbox{$\rm \nabla [Fe/H](\tau)$}) and the metallicity range of coeval stars (\mbox{$\text{Range}\widetilde{\mbox{$\rm [Fe/H]$}}(age)$}), based on two suites of cosmological simulations. Assuming a linear metallicity gradient, the birth radii for a star is expressed as:
\begin{equation} \label{eqn:rb}
\mbox{$\rm \text{R}_\text{b}$}(\text{age}, \mbox{$\rm [Fe/H]$}) = \frac{\mbox{$\rm [Fe/H]$} - \mbox{$\rm [Fe/H](0, \tau$)}}{\mbox{$\rm \nabla [Fe/H](\tau)$}}.
\end{equation}
where \mbox{$\rm [Fe/H](0, \tau$)} and \mbox{$\rm \nabla [Fe/H](\tau)$} are interpolated from Table A1 of Lu et al. 2022 \cite{2022arXiv221204515L}.

Despite potential disruptions of Interstellar Medium (ISM) metallicity caused by merger events, the calculation of birth radii remains robust. It is found that birth radii can still be inferred even during such events, as the disruption of ISM metallicity is temporary and settles within 200 Myr \cite{2022MNRAS.515L..34L}, allowing for integration into the Galaxy's ISM. Furthermore, recovery tests for galaxies that experienced mergers in the NIHAO (Numerical Investigation of a Hundred Astronomical Objects) \cite{2024MNRAS.532..411L} and TNG50 \cite{2024A&A...690A.352R} simulations show that birth radii can be accurately reconstructed, even during periods of enhanced star formation triggered by major mergers. This supports the validity of using birth radii to study radial migration, even in the presence of complex dynamical processes.

\subsection*{The Simulations}\label{simulation}

The galaxy simulation employed in this work originates from the NIHAO-UHD project \cite{Buck2018,Buck2020b}, a suite of high-resolution cosmological hydrodynamical simulations of MW-mass galaxies. The model was generated using the GASOLINE2 smoothed particle hydrodynamics solver \cite{Wadsley2017}, with cosmological initial conditions and physical processes, including star formation and feedback, implemented following the methods of Stinson et al. 2006 \cite{Stinson2006} and Stinson et al. 2013 \cite{Stinson2013}. The simulated galaxy has a total stellar mass of $1.59 \times 10^{11} M_\odot$ and is resolved with $8.2\times10^6$ star particles, $2.2\times10^6$ gas particles, and $5.4\times10^6$ dark matter particles, achieving a baryonic mass resolution of $3\times 10^4 M_\odot$ per star particle (approximately $9 \times 10^4 M_\odot$ gas particle mass) and a force softening length of 265 pc.

For this study, we first scaled the simulation to the MW scale for better comparison. We do this by multiplying all the position values ($x$, $y$, $z$, R$_{\rm birth}$) and the velocity values ($v_x$, $v_y$, $v_z$) in the simulation by the ratio of the MW scale length to the simulation scale length, which is 3.5/5.6, and the MW orbital velocity to the simulation orbital velocity after the velocity curve becomes flat, which is 240/340, respectively. To mimic the spacial extent of the data, we selected star particles that are within 7-9 kpc with $|$Z$_{\rm Gal}$$|$ $<$ 0.4 kpc.

\subsection*{Bayesian Linear Fits to the Radial Abundance Profiles}\label{plot}

In Fig.\ref{fig:rg_feh_delta_fe} and Fig.\ref{fig:rz_o_evo_radial}, we shown the results of Bayesian fits to the radial [Fe/H]/[O/Fe] abundance distributions in age bins of 1 Gyr, using the Bayesian fitting code from Anders et al. 2023 \cite{2023arXiv230408276A}(\url{https://github.com/fjaellet/xgboost_chem_ages}). We present the detailed results of these fits for age bins of 2-3 Gyr in Supplementary Figure 3 and Supplementary Figure 4.


\subsection*{The Oxygen Abundance of Young Stars}\label{oxygen}

To verify that our findings are not caused by artefacts due to selection effects, we plot the Kiel diagram of the stars from the GALAH DR3 in Supplementary Figure 5a, color-coded by oxygen abundance. As shown in the Supplementary Figure 5, the high temperature MSTO stars with 6200 K $<$ $T_{\rm eff}$ $<$ 7000 K behave oxygen-enhancement compared with stars at lower temperature end. Most of these oxygen-enhanced stars at high temperature end have ages less than 4 Gyr, which is consistent with our result in Section Age–abundance distribution of the Milky Way disc. Consequently, the oxygen-enhancement in young disc stars in Fig.\ref{fig:rg_feh} is not due to selection effects, but is directly observed by GALAH survey, and the precise ages of our sample stars allow us to accurately characterise the variation in oxygen abundance of disc stars. In addition, we have examine the oxygen abundance of about 15000 common stars from GALAH DR3 and APOGEE DR17, finding no significant systematic differences in oxygen abundances up to 7000 K.

\subsection*{Impact of the Selection Function of GALAH}\label{sel galah}

Sahlholdt et al. (2022) \cite{2022MNRAS.510.4669S} has discussed the impact of the GALAH sample's selection function, which enhances the young peak at an age of 3 Gyr. Since young stars are brighter on average than older stars, the magnitude limit of GALAH can skew the age distribution towards the young end, especially for the distant populations. However, in our study, we divided the sample stars into spatial bins based on R$_{\rm guide}$ and z$_{\rm max}$. The key findings focus on the age-abundance relations of the local disc at low-z$_{\rm max}$ region, which mitigates selection effects from the GALAH survey's magnitude limit.

\end{appendices}



\section*{Data availability}


We made use of publicly available data in this work. GALAH data are available at \url{https://cloud.datacentral.org.au/teamdata/GALAH/public/GALAH_DR3/}, Gaia data at \url{https://cdn.gea.esac.esa.int/Gaia/gdr3/}. The simulation data used in this work are available at \url{http://tobias-buck.de/#sim_data}. The luminosities of sample stars from Yu et al. 2023 \cite{2023ApJS..264...41Y} are available at \url{https://cdsarc.cds.unistra.fr/viz-bin/cat/J/ApJS/264/41#/browse}.
This work has made use of data from Anders et al. 2023 (\url{https://github.com/fjaellet/xgboost_chem_ages}) \cite{2023arXiv230408276A}
The stellar ages, guiding radii, and birth radii of sample stars are available at \url{https://zenodo.org/records/14326332}. Source data are provided with this paper, including all data used to produce Figs. 1–10 and Supplementary Figures 1–5. The datasets generated during and/or analysed during the current study are available from the corresponding author upon request.


\section*{Code availability}

This work is made possible by the following open-source software: NumPy \cite{2011CSE....13b..22V}, SciPy \cite{2020NatMe..17..261V}, Matplotlib \cite{2007CSE.....9...90H}, Astropy \cite{2013A&A...558A..33A,2018AJ....156..123A}, Galpy \cite{2015ApJS..216...29B}, Seaborn\cite{Waskom2021}, Pandas \cite{mckinney2010data}, Emcee \cite{2013ascl.soft03002F}, and Statsmodels \cite{seabold2010statsmodels}.
This work has made use of analysis code from Anders et al. 2023 \cite{2023arXiv230408276A}(\url{https://github.com/fjaellet/xgboost_chem_ages}).
The GASOLINE2 code is publicly available at \url{https://gasoline-code.com}, with the modified version used in this study is hosted at \url{https://github.com/N-BodyShop/gasoline}.
No new codes are developed in this paper.

\section*{Acknowledgements}

The authors thank Joss Bland-Hawthorn and Sven Buder for their discussions on the reliability of the oxygen abundances in the star sample. The authors acknowledge Thomas G. Bisbas for valuable assistance in refining the manuscript's language.
%
%
This work is supported by the NSFC grant (12090040, 12090042) (S.B.), the Joint Research Fund in Astronomy (U2031203) under cooperative agreement between the National Natural Science Foundation of China (NSFC) and Chinese Academy of Sciences (CAS)(S.B.), the NSFC grant 12403037 (X.C.), the NSFC grant 12373020 (Y.C.), the National Key R$\&$D Program of China No. 2023YFE0107800 (Y.C.), the NSFC grant 12473028 (X.Z.), the NSFC grant 12303035 (Z.G.), and the Beijing Natural Science Foundation 1242016 (Z.G.).

\section*{Author Contributions Statement}
 

T.S. contributed stellar age determination and wrote the initial manuscript. S.B. supervised T.S. on the stellar modelling. T.S., S.B. and X.C. provided the ideas to initialize the project and conduct all aspects of the analysis. 
Y.C. contributed to discussions on the age-[Fe/H] relation and the division of sample stars.
Y.Lu ran the calculations to derive birth radius and helped write the manuscript.
C.L. helped with explored the impact of the selection effects.
Y.Lu and T.B. provided the cosmological hydrodynamical simulations and supported the data analysis.
X.Z. and T.L. helped with discussions on the chemical abundance.
Y.C., C.L., Y.Li, Y.W., Z.G. and L.Y. improved the text.
All authors discussed and commented on the manuscript.

\section*{Competing Interests Statement}

The authors declare no competing interests.



\end{document}